\documentclass[11pt,a4paper]{article}
\pdfoutput=1
\usepackage{jcappub}
\newif\ifonecol\onecolfalse
\usepackage{amsfonts}
\usepackage[textsize=scriptsize,backgroundcolor=red!70,linecolor=red]{todonotes}
\usepackage{array}
\usepackage{slashed} 
\usepackage{dcolumn} 
\usepackage{paralist}
\usepackage{mathrsfs} 
\usepackage{todonotes}
\usepackage{subcaption}

\newcommand{\nue}{\ensuremath{\nu_e}}
\newcommand{\numu}{\ensuremath{\nu_\mu}}
\newcommand{\nutau}{\ensuremath{\nu_\tau}}

\newcommand{\ie}{\textit{i.e.}}
\newcommand{\eg}{\textit{e.g.}}

\newcommand{\anti}[1]{\ensuremath{\bar{#1}}}

\newcommand{\dm}{DM}

\newcommand{\pdm}{\ensuremath{\text{PDM}}}
\newcommand{\fdm}{\ensuremath{\text{FDM}}}

\newcommand{\ord}[1]{\ensuremath{\mathcal{O}(#1)}}
\newcommand{\zp}{\ensuremath{Z^{\prime}}}

\newcommand{\gccz}{\ensuremath{g_{\chi\chi Z}}}
\newcommand{\fgal}{\ensuremath{\Phi^\text{G}}}
\newcommand{\fxgal}{\ensuremath{\Phi^\text{EG}}}
\newcommand{\ud}{\ensuremath{\mathrm{d}}}
\newcommand{\fdnde}{\ensuremath{\frac{\ud N_\chi}{\ud E_{\chi}}}}

\newcommand{\cmssr}{\ensuremath{\text{ cm}^{-2}\text{ s}^{-1}\text{ sr}^{-1}}}

\newcommand{\intg}[3]{\ensuremath{\int_{#1}^{#2}\!\!\!\ud {#3}\,}}
\newcommand{\Ein}{\ensuremath{E^\text{in}_{\chi}}}
\newcommand{\Edep}{\ensuremath{E^\text{dep}}}

\title{The Direct Detection of Boosted Dark Matter at High Energies and PeV events at IceCube}

\author[a]{Atri Bhattacharya,}
\author[b,c]{Raj Gandhi}
\author[b]{and Aritra Gupta}
\affiliation[a]{Dept.\ of Physics, University of Arizona,\\1118 E.\ Fourth Street, Tucson, AZ 85721, United States of America}
\affiliation[b]{Harish-Chandra Research Institute,\\Chhatnag Road, Jhunsi, Allahabad-211019, India}
\affiliation[c]{Fermi National Accelerator Laboratory,\\P.O. Box 500, Batavia, IL 60510, United States of America}

\emailAdd{atrib@email.arizona.edu}
\emailAdd{nubarnu@gmail.com}
\emailAdd{aritra@hri.res.in}

\abstract{
We study the possibility of detecting dark matter directly via a small but  energetic component that is allowed within present-day constraints.
Drawing closely upon the fact that neutral current neutrino nucleon interactions are indistinguishable from DM-nucleon interactions at low energies,  we extend this feature to high energies for a small, non-thermal but highly energetic  population of DM particle $\chi$, created via the decay  
of a significantly more massive and long-lived non-thermal relic $\phi$, which forms the bulk of DM.
If $\chi$ interacts with nucleons, its cross-section, like the neutrino-nucleus coherent cross-section, can rise sharply with energy leading to deep inelastic scattering, similar to neutral current neutrino-nucleon interactions at high energies.
Thus, its direct detection may be possible via cascades in very large neutrino detectors.
As a specific example, we apply this notion to the recently reported three ultra-high energy PeV cascade events clustered around $1-2$ PeV at IceCube (IC). We  discuss the  features which may help discriminate this scenario from  one in which only astrophysical neutrinos constitute the  event sample in detectors like IC.
}

\arxivnumber{1407.3280}


\begin{document}

\maketitle

\section{Introduction}
The nature and origin of dark matter (DM) remains one of the principal unanswered questions in physics. While theoretical biases have served as a guide for searches and model-building,  in principle, very little is known about its nature and properties. Specifically, the DM mass can span the range $10^{-15}$--$10^{15}$ GeV, and its interaction cross-section with nucleons and  annihilation cross-section into SM particles can lie in the range $10^{-76}$--$10^{-41}$ cm$^2$ \cite{Ibarra}. 

Since the bulk of DM is known to be non-relativistic, its direct detection has focussed on its low-energy coherent scattering off nuclei, leading to nuclear recoils which have energies of a few keV, making them very challenging to detect over backgrounds. 
In general, most efforts have been directed towards exploring  the parameter space spanned by thermal DM masses in the 10--100 GeV range with weak-scale interaction cross-sections with nucleons.
Recent experiments have, however, significantly constrained this space for such particles (also called WIMPS).\footnote{For recent reviews, see \cite{1406.5200, Ibarra}.}
When combined with results from indirect DM searches and colliders, it is fair to say that credible reasons for seriously considering ``non-WIMP'' and possibly non-thermal candidates for DM exist (for a review, see \cite{1310.8642}). 
In addition, we note that Big Bang Nucleosynthesis (BBN) constraints derived from the primordial Helium and Deuterium abundances \cite{Nollett:2014lwa}, indicate that the number of effective relativistic species is $N_{eff}= 3.56\pm0.23$. Constraints derived from observations of the Cosmic Microwave Background (CMB) by the Planck experiment \cite{Ade:2013zuv} similarly favor the presence of some ``dark radiation'' over and above the three standard model neutrinos, with $N_{eff}= 3.30\pm0.27$. Significantly, when combined, these two sets of constraints with very different origins rule out the presence of a full sterile neutrino, $\Delta N_{eff}=1 \text{ at } > 99\%$ C.L, whereas the absence of any additional neutrino coupled relativistic species is disfavoured at $> 98\% C.L.$ \cite{Nollett:2014lwa}. Relativistic non-thermal DM particles could be one possible way of resolving this \cite{Hooper:2011aj}.

The possibility that DM  may be a multi-particle sector has, of course, been extensively studied in the literature under various assumptions. Due to the reasons mentioned above, it is possible that  the bulk of this  sector may comprise of non-thermal  (and non-relativistic) components, and may  also  contain  a small component that may be relativistic and highly energetic. In what follows, we focus on the detection of DM via this component. Specifically, we explore the possibility of directly detecting  DM  in existing large neutrino detectors at energies much higher than presently considered.  After  further motivating  this idea,  we  explore its consequences qualitatively and quantitatively when specifically applied to the recently announced IceCube (IC) PeV events.

We  note that coherent elastic neutrino-nucleus scattering \cite{Freedman:1973yd}, a process not yet experimentally observed due to the very small nuclear recoil measurement required to detect it, is expected to be an irreducible background for future DM direct detection experiments \cite{Billard:2013qya}. Thus DM and neutral current (NC) neutrino interactions mimic each other at low energies. One can expect that this analogy holds with rising energies, and in particular at the highest energies at which neutrinos are presently detected.

In the present work, we assume DM to be primarily non-thermal\footnote{Unitarity bounds constrain  particles with mass $m \gtrsim 300$ TeV to remain out of thermal equilibrium throughout its history as discussed in \cite{PhysRevLett.64.615}; in addition, we choose the lighter \dm\ species to be non-thermal also.} with its bulk comprised of a very massive relic $\phi$ (with mass $m_\phi$ and a lifetime $\tau_\phi$ greater than the age of the Universe) which decays preferentially to another much lighter DM particle $\chi$ (as opposed to decaying to standard model (SM) daughters). This leads to a small but significant population of ultra-high energy relativistic DM particles, non-thermally created in the narrow energy region spanning $m_\phi$.

Drawing closely upon the similarity between neutrino NC and DM interactions, we further assume that $\chi$ interacts with SM particles with cross-sections much smaller than standard weak interactions via the exchange of a heavy gauge boson which connects the SM and DM sectors. The assumption of a small strength  interaction between DM and SM is of course empirically required, and the assumption about the existence of a heavy neutral guage boson provides  a simple way to implement it.
At high energies, this will result in deeply inelastic interactions (DIS) of DM with SM particles, and mimic UHE neutrino-nucleon NC interactions \cite{Gandhi:1995tf,Gandhi:1998ri} in a detector like IC, creating cascades which are indistinguishable from those created by neutrinos.

In what follows, we quantitatively implement the above proposal of looking for DM at high energies in neutrino detectors  by performing a  flux and cross-section calculation. While the approach is generic, it can be modified in specific ways to perform a broader and more general study, vis a vis choices of a mediator (scalar versus a vector boson, for instance), coupling strengths , masses etc \cite{ab_in_progress}. Our choices below are pertinent to our chosen application, which are the recently observed PeV IC events.
We  calculate the DM-nucleon cross section at high energies in analogy with the neutrino-nucleon NC cross-section. We then focus on the three PeV events in IC, and assuming that their cascades originate in DM interactions with ice nuclei, we determine the ramifications for DM mass and flux which result from this. Finally, we discuss the general  features that would distinguish this scenario from others in which all the events in IC -like detectors are due to neutrino scattering.

\section{Neutral-current scattering of a relativistic dark matter species with nuclei}
 We assume that the \dm\ sector consists of at least two particle species with the following  properties:
\begin{itemize}
	\item A co-moving non-relativistic  real scalar species $\phi$,  with a mass of \ord{10 \text{ PeV}}, which is unstable but decays  with a very large lifetime to $\chi$, and does not have any decay channels to SM particles. We call this species the PeV Dark Matter (\pdm), and it comprises the bulk of present-day DM.
	\item A lighter fermionic \dm\ species (FDM), $\chi$ with mass $m_\chi \ll  m_\phi$, which we assume is produced in a monochromatic pair when the \pdm\ decays, \ie, $\phi \to \bar{\chi}\chi$,   each with energies of  $m_{\phi} / 2$.\footnote{As mentioned above, the choice of a PeV scale mass for DM and subsequent choices of couplings and a mediator is based on our application below  to recent IC events, but they are representative of a concept that may have broader applicability.}
\end{itemize}

The lifetimes for the decay of heavy DM particles to standard model species are strongly constrained ($ \tau \gtrsim 10^{27} $--$ 10^{28} $s) by diffuse gamma-ray and neutrino observations \cite{Murase:2012xs,Rott:2014kfa}.
However, since in our scenario $\phi$ does not decay to SM particles, constraints relevant here are only those based on cosmology, which limits the total relativistic particle density of the universe at the respective epochs, independent of what those particles are, and are significantly weaker.
Specifically, these include limits from the observed CMB anisotropies \cite{astro-ph/0403164}, light nuclei abundances during Big-Bang Nucleosynthesis (BBN) \cite{Dev:2013yza, Nollett:2014lwa} and from structure formation (see, \eg, \cite{DelPopolo:2008mr} for a review).\footnote{BBN is also sensitive to the electron-positron pair production rate in \dm\ annihilation, but for both the \pdm\ and \fdm\ these interaction strengths are tiny.}
Consistent with these constraints, and with present-day relic abundance considerations, we assume that the \pdm\ decays with a lifetime of $\tau_{\phi} \gtrsim 10^{17}$s, \ie, greater than the lifetime of the universe.
Additionally,  the lighter (and stable) \fdm\ species is assumed to be produced only non-thermally, via the decay of the long-lived  \pdm. 
Its contribution to the  DM mass density is thus expected to be small.

The \fdm\ flux is composed of galactic and extragalactic components of comparable magnitudes \cite{Esmaili:2012us}.
Thus, the total flux $\Phi = \fgal + \fxgal$, where, $\fgal$ and $\fxgal$ respectively represent the galactic and extra-galactic components of this flux \cite{Esmaili:2012us, 1311.5864}):
\begin{equation}\label{eqn:difffgal}
\fgal = \intg{E_\text{min}}{E_\text{max}}{E_\chi} D_\text{G} \fdnde,
\end{equation}
and,
\begin{subequations}\label{eqn:difffxgalall}
\ifonecol
\begin{align}
\fxgal &= \frac{\Omega_{\dm}\,\rho_{\text{c}}}{4\pi\,m_{\phi}\,\tau_{\phi}}
\intg{E_\text{min}}{E_\text{max}}{E_\chi}\! \intg{0}{\infty}{z} \frac{1}{H(z)} \fdnde \left[(1+z)E_{\\chi}\right]\\
\label{eqn:difffxgal}
&= D_\text{EG} \intg{E_\text{min}}{E_\text{max}}{E_\chi}\! \intg{0}{\infty}{z} \frac{1}{\sqrt{\Omega_{\Lambda} + \Omega_\text{m}(1+z)^{3}}} \fdnde \left[(1+z)E_{\chi}\right],
\end{align}
\else
\begin{align}
\fxgal &= \frac{\Omega_{\dm}\,\rho_{\text{c}}}{4\pi\,m_{\phi}\,\tau_{\phi}}
\intg{E_\text{min}}{E_\text{max}}{E_\chi} \!\! \intg{0}{\infty}{z} \frac{1}{H(z)} \fdnde \left[(1+z)E_{\chi}\right]\\
\label{eqn:difffxgal}
&= D_\text{EG} \intg{E_\text{min}}{E_\text{max}}{E_\chi} \!\!\intg{0}{\infty}{z} \frac{1}{\sqrt{\Omega_{\Lambda} + \Omega_\text{m}(1+z)^{3}}}\nonumber\\
&\quad\quad\quad\quad\quad \times\fdnde \left[(1+z)E_{\chi}\right],
\end{align}
\fi
\end{subequations}
with
\[D_\text{G} = 1.7\times 10^{-8} \left( \frac{1\text{ TeV}}{m_{\phi}} \right)
\left( \frac{10^{26}\text{ s}}{\tau_{\phi}} \right) \cmssr\]
and
\[D_\text{EG} = 1.4 \times 10^{-8} \left( \frac{1\text{ TeV}}{m_{\phi}} \right)
\left( \frac{10^{26}\text{ s}}{\tau_{\phi}} \right) \cmssr.\]
Here, $z$ represents the red-shift of the source, $\rho_\text{c} = 5.6\times 10^{-6}\text{ GeV cm}^{-3}$ denotes the critical density of the universe, and we have used $H(z) = H_{0}\sqrt{\Omega_{\Lambda} + \Omega_\text{m}(1+z)^{3}}$, and $\Omega_\Lambda = 0.6825$, $\Omega_\text{m} = 0.3175$, $\Omega_{\dm} = 0.2685$ and $H_{0} = 67.1 \text{ km}\text{ s}^{-1}\text{ Mpc}^{-1}$ from the recent PLANCK data \cite{Ade:2013zuv}.
For the two-body decay $\phi \to \anti{\chi}\chi$
\begin{equation}
\fdnde = 2\delta\left(E_\chi - \frac{1}{2}m_{\phi}\right)\,,
\end{equation}
where, $E_\chi$ denotes the energy of each of the produced $\chi$ particle.

\begin{figure}[htb]
	\centering
	\begin{subfigure}{0.4\textwidth}
		\centering
		\includegraphics[width=0.75\textwidth]{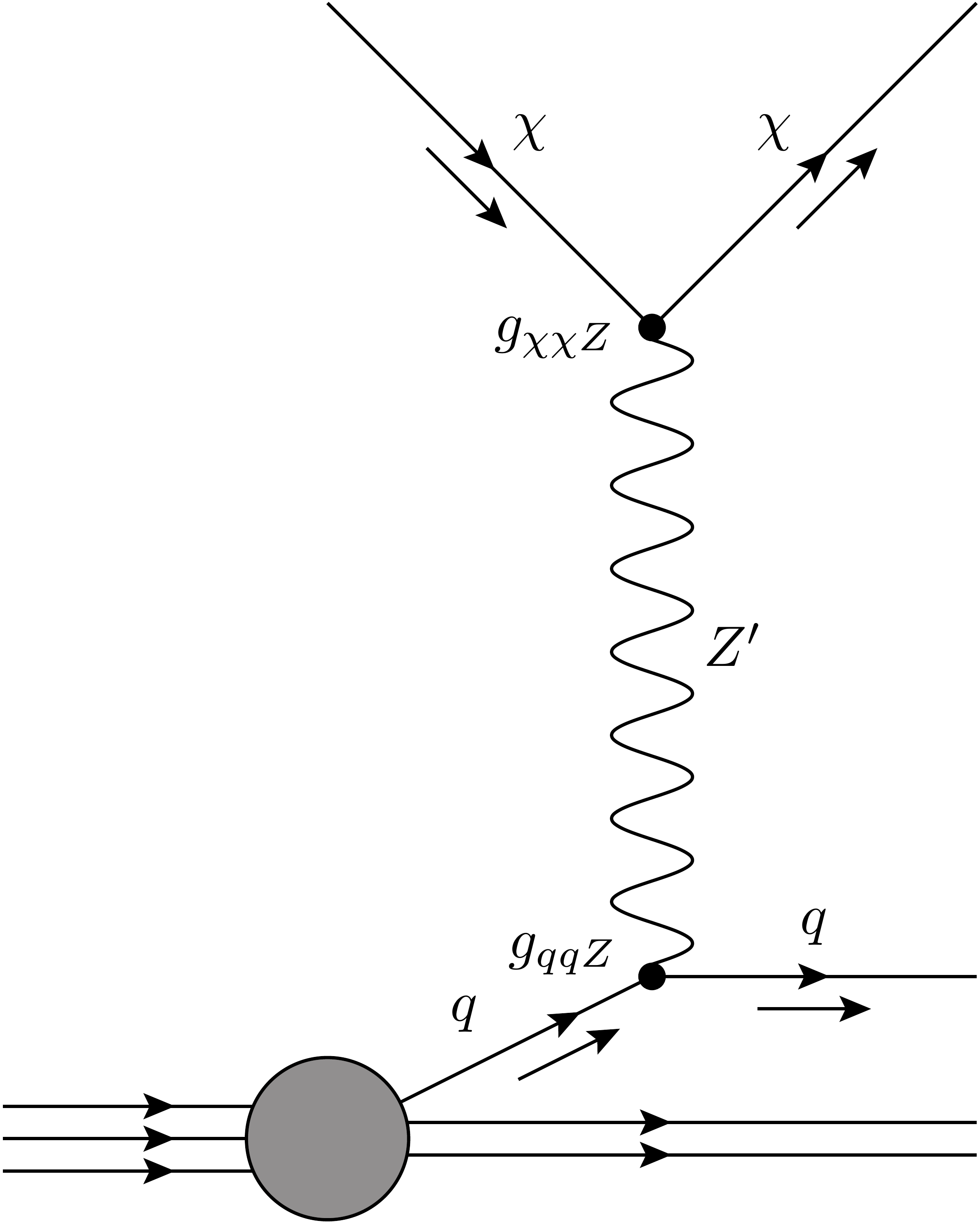}
		\caption{}
		\label{fig:tdm_nucl}
	\end{subfigure}
	\begin{subfigure}{0.59\textwidth}
		\centering
		\includegraphics[width=\textwidth]{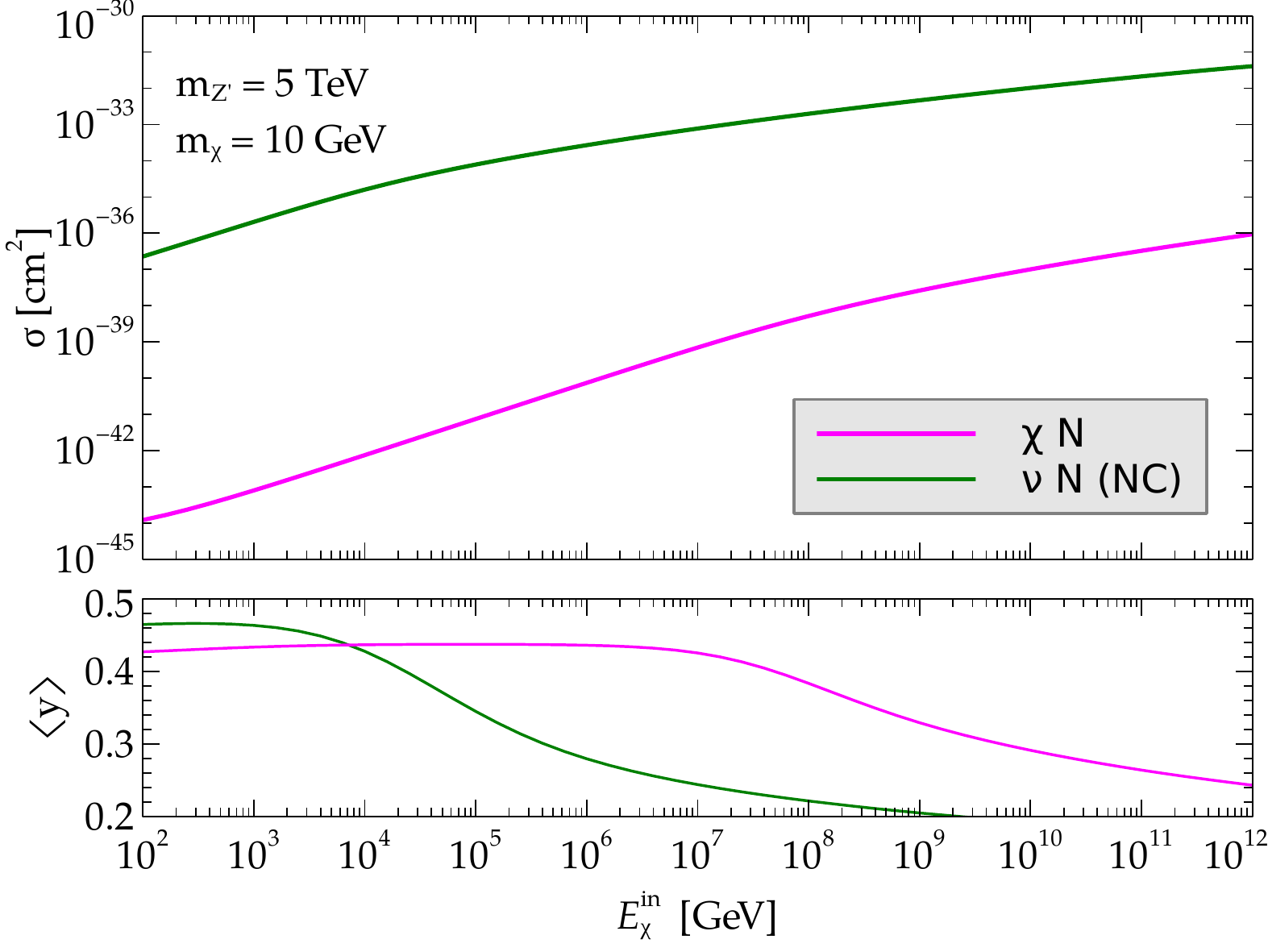}
		\caption{}
		\label{fig:dmsigma}
	\end{subfigure}
		\caption{(a) Interaction of the incoming TeV mass \dm\ particle $\chi$ with a nucleus, mediated by a heavy non-standard boson \zp. (b) The ${\chi}N$ DIS interaction cross-section and the corresponding $\langle y(E) \rangle$ are shown for the benchmark value of $m_\chi$ and $m_{Z^\prime}$. The overall normalisation to the $ \chi N $ cross-section is set by the product of coupling constants $G$, and is here arbitrarily chosen to be $G = 0.05$. The real magnitude of $G$ will be determined by comparing event rates to those seen at IC in the succeeding section.
		For comparison, the $ \nu N $ neutral current cross-section and the corresponding $ \langle y\rangle $ are also shown.}
\end{figure}

The \fdm\ interacts with the nucleus within the IceCube detector via a neutral current interaction mediated by a beyond-SM heavy gauge boson, \zp\ (Fig.\ \ref{fig:tdm_nucl}) that couples to both the $\chi$ and quarks and gluons.

For both the $\chi\chi\zp$ and $qq\zp$ interactions we assume the interaction vertex to be vector-like, with hitherto undetermined coupling constants \gccz\ and $g_{qqZ}$ respectively.\footnote{We have deliberately tried to avoid limiting the scenario to any particular theoretical model in order to focus solely on the phenomenological signatures of the two-sector DM that we have discussed here.
Theoretical models that encompass our DM spectrum have been discussed in the literature in terms of $Z$ or $ Z^{\prime} $ portal sectors with the $ Z^{\prime} $ vector boson typically acquiring mass through the breaking of an additional $ U(1) $ gauge group at the
high energies (see \eg, \cite{Alves:2013tqa,Hooper:2014fda}).}
The DIS cross-section for $\chi N \to \chi X$ is then computed in the lab-frame, with the product $G = \gccz g_{qqZ}$ as the undetermined parameter, over a broad range of incoming \fdm\ energies, $100\text{ GeV} \leq E^\text{in}_{\chi} \leq 10 \text{ PeV}$, using tree-level \verb+CT10+ parton distribution functions \cite{Lai:2010vv}.
We set the \zp\ mass to be 5 TeV.
For \zp\ with mass $\geqslant 2.9$ TeV, the couplings \gccz\ and $g_{qqZ}$ are largely unconstrained by collider searches \cite{Buckley:2011vc}, thus are limited only by unitarity.\footnote{We note here that due to the presence of $ \chi\chi \zp $ vertex, the possibility that \zp-bremsstrahlung affects the two-body $ \phi\to \chi\chi $ decay and thus the energies of the outgoing $ \chi $-particles becomes worth considering.
We have verified by means of explicit calculations that, for the value of the parameters $ G^2 $ and $ \tau_\phi $ that we require in order to fit the predicted events from $ \chi N $ NC scattering with IC observations (see section \ref{dmfit}), \zp\ bremsstrahlung-included decay rate is about 5\% of the total decay rate and therefore negligible. A presentation of the full computation is beyond the scope of this paper, but closely follows a similar computation made in \cite{Kachelriess:2009zy}.}

Since the IC can only measure the deposited energy \Edep\ for neutral current events, it is important to determine the nature of the inelasticity parameter, relating the deposited energy to the incoming particle energy ($E^\text{in}_\chi$):
\begin{equation}\label{eq:ydef}
y = \frac{\Ein - E^\text{out}_{\chi}}{E^\text{in}_{\chi}} = \frac{\Edep}{E^\text{in}_{\chi}}\,,
\end{equation}
where, $E_\chi^\text{out}$ represents the energy of the outgoing $\chi$ in the scattering process.
The DIS differential cross-section with respect to the inelasticity parameter is then expressed as
\begin{equation}\label{eq:diffsig}
\frac{\ud \sigma}{\ud y}(\Ein, y) = G^{2} f(\Ein, y)\,.
\end{equation}

The results for the total cross-section and the mean inelasticity parameter,
\[
\langle y(E^\text{in}_\chi)\rangle = \frac{1}{\sigma(E^\text{in}_\chi)}\int_0^1\!{\ud y}\,y\frac{\ud \sigma(E^\text{in}_\chi, y)}{\ud y}\,,
\]
are shown in Fig.\ \ref{fig:dmsigma}.

\section{The IC events: Characteristics and possible Origins}

Prior to applying our proposal, we recapitualte the basic observations and features of the IC data below.

The  observation of ultra-high energy (UHE, $E_\nu\geq 30$ TeV)  neutrino events  at IceCube (IC) \cite{Aartsen:2013jdh, Aartsen:2014gkd} is one of the most striking of recent  experimental results in all of physics. When statistically buttressed by  imminent additional observations by IC and other high energy neutrino observatories like ANTARES \cite{Adrian-Martinez:2014wzf}, AUGER \cite{Scherini:2014gja} and the upcoming KM3NET \cite{Margiotta:2014eaa} they promise to open hitherto unprecedented windows of understanding on the highest energy processes in our Universe.

In IC, neutrino detection occurs via weak charge and neutral current (CC and NC respectively) interactions with nucleons in ice, resulting in the deposition of visible energy in the form of Cerenkov radiation. Observed events are categorized into two distinct  types:
\begin{itemize}
	\item \numu\ CC and a subset of \nutau\ CC interactions produce \emph{tracks} of highly energetic charged leptons traversing a significant length of the detector, while
	\item \nue\ CC, a subset of \nutau\ CC and NC interactions of all three flavors produce \emph{cascades} characterized by their collective light deposition in a bulbous signature distributed around the interaction vertex.
\end{itemize} 
Additionally, in spite of the belief that sources do not produce  $\nu_\tau$, the flavour ratios for neutrinos  are rendered close to $1:1:1$ at earth due to  oscillations over large distance scales.
In this situation, cascade events are expected to  constitute about 75--80\% of the total observed sample \cite{Beacom:2004jb}. The background to these events is provided by the rapidly falling atmospheric neutrino flux and the muons created in cosmic-ray showers in the atmosphere.

The 988-day IC data reveals 37 events (9 track, 28 cascades) with energies between $30$ TeV and $2$ PeV, consistent with a diffuse neutrino flux given by
\begin{equation}\label{eqn:ic-bf}
E^2\phi_\nu(E) = 0.95 \pm 0.3 \times 10^{-8}\text{ GeV cm}^{-2}\text{ s}^{-1}\text{ sr}^{-1},
\end{equation}
in the energy range 60 TeV--3 PeV, where $\phi_\nu$ represents the per-flavor flux.
A purely atmospheric/cosmic-ray shower origin of these events is rejected at the $5.7\sigma$ level.

We mention three characteristics of the event sample which will be pertinent to our work below:
\begin{inparaenum}[\itshape a\upshape)]
	\item the three highest energy events are closely clustered, with  energies of 1 PeV, 1.1 PeV and  2.1 PeV,
	\item there are no events between 400 TeV and 1 PeV, a gap which can be statistically realised in  43\% of continuous power-law spectrum predictions \cite{Aartsen:2013jdh, Aartsen:2014gkd} , and
	\item there are no events beyond 2 PeV, although 3 events are expected between 3--10 PeV for an unbroken $E^{-2}$ spectrum.\footnote{This expectation is  due to the Glashow resonance \cite{Glashow:1960zz, Bhattacharya:2011qu, Barger:2014iua}.}
\end{inparaenum}

The precise origin of these events is as yet unknown. There is weak evidence of a slight Galactic bias in the directionality, but the overall distribution over the entire sample is consistent with a diffuse isotropic flux.
Possible astrophysical sources including both from within our galaxy \cite{1403.3206,1309.4077,1309.2756,1311.7188,1405.3797,1305.6606,1310.7194} and from outside the galaxy \cite{astro-ph/9609048,astro-ph/0601695,1306.3417,1303.1253,PhysRevLett.66.2697, 1305.7404,astro-ph/9701231, 1306.2274} have been considered as explanations for the origin of these high energy particles.
Some models of astrophysical sources, \eg, for galaxy clusters \cite{Murase:2008yt} and starburst galaxies \cite{Loeb:2006tw}, also predict a break in the neutrino spectrum at energies above $ \sim 1 $ PeV consistent with IC observations.
In addition, the possibility that such UHE events might originate from the decay/annihilation of super-heavy DM into standard model particles has also been investigated \cite{1303.7320, 1308.1105, 1311.5864, Zavala:2014dla, Bhattacharya:2014vwa}.

\subsection{\label{dmfit}PeV events: Fitting the DM-prediction to the IC observation}
We next determine the values of the parameters $ G^2 $ and $ \tau_\phi $ that fits the number of DM events from our prediction with the IC PeV events.
The energy at which the $\chi$ flux should peak is determined by requiring that the event rates peak at around 1.1 PeV; in turn, this requires that the flux peak at around energies of
\[E_\text{peak} = 1.1 / \left[\langle y\rangle\big\vert_{E^\text{in}_\chi = 1.1\text{ PeV}}\right] = 2.53\text{  PeV},\] which implies, $m_{\phi} = 5.06$ PeV.

The total number of events in a given IC bin increases proportionally with the incident flux and the interaction rate of the incident particles with the ice nuclei relevant to the corresponding bin energies.
Since, in addition, the \fdm\ flux $\Phi \propto \tau_\phi^{-1}$ [Eq.\ \eqref{eqn:difffgal} and \eqref{eqn:difffxgalall}] and $\ud\sigma/\ud y \propto G^{2}$ [Eq.\ \eqref{eq:diffsig}], the ratio $G^2 / \tau_\phi$ of the undetermined parameters G and $\tau_\phi$ can be ascertained by normalising the number of events predicted due to the \fdm\ flux at deposited energies $\Edep\geqslant 1$ PeV against those seen at the IC.
We find that for a reasonable decay lifetime of $\tau_\phi = 5\times10^{21}$s, we need to set $G = 0.047$ to obtain the 3 PeV+ events from the \fdm\ flux seen over the 988-day IC runtime.
The values of the parameters thus determined are well within the allowed parameter-space, given constraints on the coupling constant from perturbativity and on the lifetime from model independent considerations for heavy DM decaying to relativistic particles: $\tau_{\phi} \geq10^{18}$s \cite{Audren:2014bca}.
The corresponding nature of the \fdm\ extragalactic flux is shown in Fig.\ \ref{fig:allflux}.
The bigger the value of $\tau_\phi$, the larger would $G$ need to be, to match the IC PeV+ event rate, with the upper bound to the coupling constant and, by consequence, the upper bound to $\tau_\phi$, being set by unitarity limits on $G$.

\subsection{Sub-PeV Events: Neutrinos from extra-galactic sources}

\begin{figure}[htb]
\centering
	\includegraphics[width=0.7\textwidth]{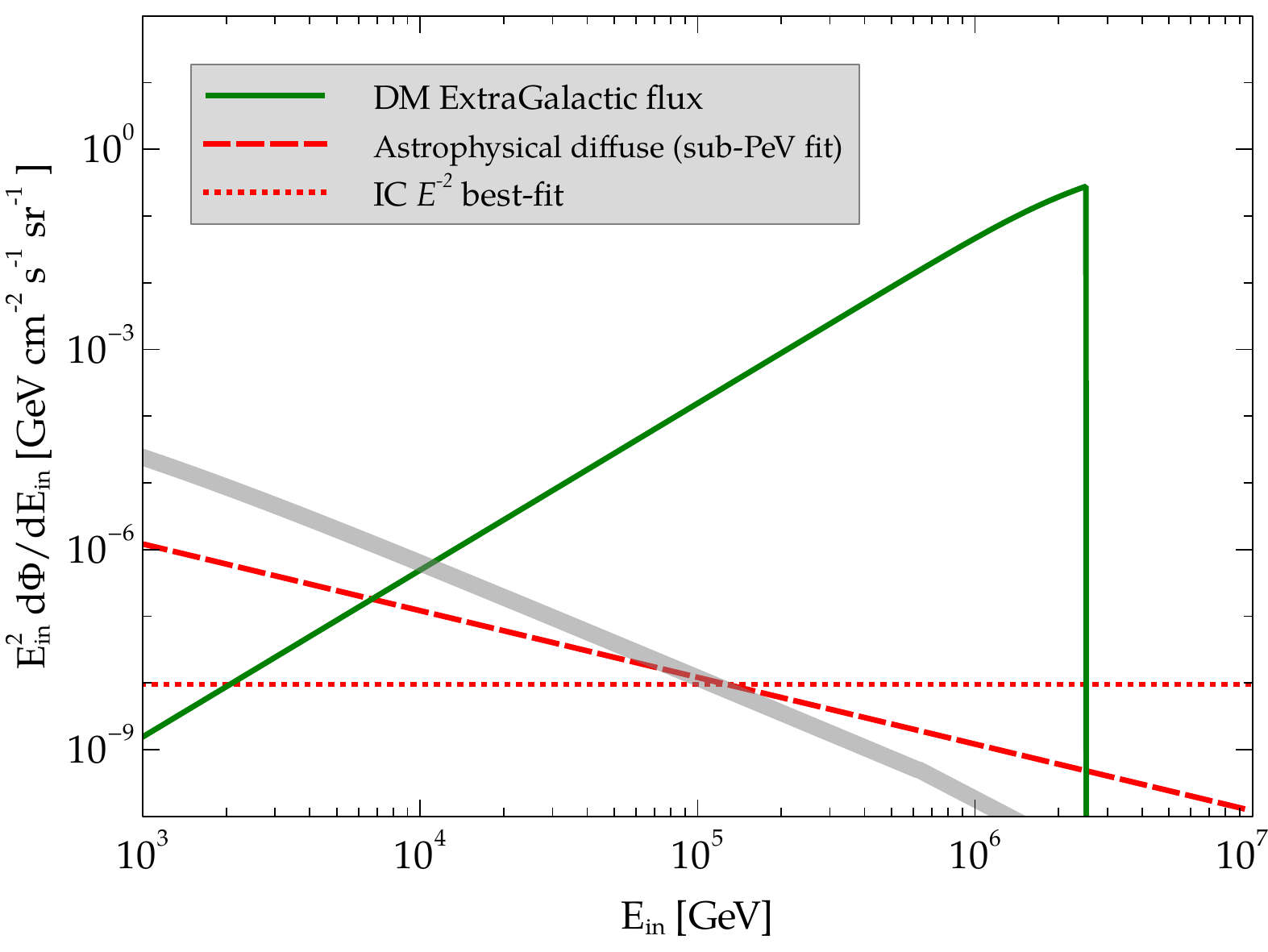}
	\caption{\label{fig:allflux}The TeV-scale diffuse neutrino flux and the extra-galactic \fdm\ flux at PeV+ energies for decay lifetime $\tau_\phi = 5\times10^{21}$s.
	The thick light-gray curve indicates the estimated conventional atmospheric $\nu_\mu + \anti{\nu}_\mu$ flux \cite{Gondolo:1995fq}.}
\end{figure}

\begin{figure}[htb]
	\centering
	\includegraphics[width=0.7\textwidth]{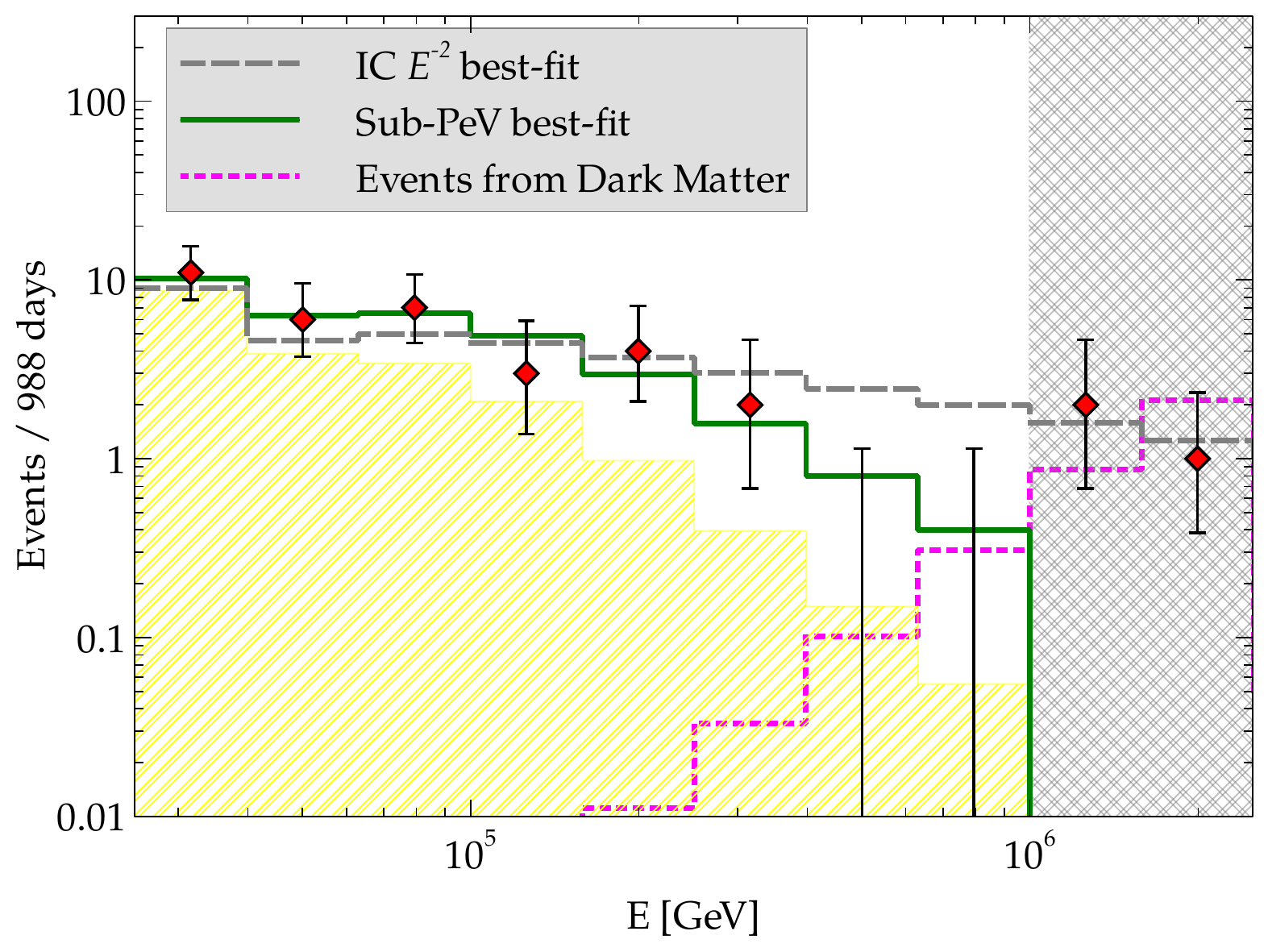}
	\caption{\label{fig:subpevfit}Predicted and observed total event rates at the IceCube. The gray shaded region represents energies at which we expect events predominantly from the \dm\ sector.
	The green line shows event-rate predictions from our best fit flux to the sub-PeV event-rates observed at IC, with the flux given by Eq.\ \eqref{eq:ourbf}.
	The event rates predicted due to the IC best-fit $E^{-2}$ flux (gray dashed line) and the observed data (red diamonds) are shown.
	The IC-estimate for the atmospheric background events is shown as the yellow shaded region.}
\end{figure}

While the events corresponding to deposited energies $E_\text{dep} \geqslant 1$ PeV are accounted for by the \fdm\ flux, the sub-PeV events up to 400 TeV are consistent with a power-law flux of incident particles, and are, likely, representative of a diffuse flux of neutrinos from extragalactic sources.
The term ``best-fit'' has limited validity at this point in time since given the limited statistics, it is at present unclear if the flux is truly diffuse and extra-galactic, or a superposition of individual extended sources or a combination of these alternatives \cite{Ahlers:2014ioa}.
Indeed, using only the sub-PeV events to determine the best-fit $E^{-\alpha}$ spectrum, we find that the IC observation is closely matched by a more steeply falling astrophysical flux spectrum than that in Eq.\ \eqref{eqn:ic-bf}, \ie, the best-fit is instead given by (Fig.\ \ref{fig:allflux}) \footnote{Theoretically we can encounter a flux spectrum that is softer than $E^{-2}$. See \eg, \cite{becker2005source}.}
\begin{equation}\label{eq:ourbf}
\mathrm{d}\Phi_\text{astro}/\mathrm{d}E_\text{in} = 1.21\times 10^{-3} E^{-3.0} \text{ GeV}^{-1}\cmssr.
\end{equation}
We note here that while Eq.~\eqref{eq:ourbf} represents the best possible fit to the sub-PeV events from a power-law,
any soft spectra with index $ \alpha \leqslant 2.5 $ and appropriate normalization would be compatible with the data,
within a $ 1\sigma $ confidence level, although with slightly poorer goodness-of-fit measures.
Due to the softness of the spectral shape, the astrophysical flux drops to below the single-event threshold at energies higher than 400 TeV, rendering it naturally consistent with the lack of events at subsequent energies up to the PeV.
The \fdm\ flux itself does not contribute appreciably to the sub-PeV event-rate (see figure \ref{fig:subpevfit}).

We note here that the gap in the event-spectrum between 400 TeV--1 PeV is not yet statistically significant and, therefore,
a diffuse astrophysical flux with less steep spectra $ \alpha \approx 2\text{--}2.3 $ would also be consistent with the sub-TeV event-spectra should this gap fill up in the future.

\section{Discussion and Conclusions}

Given present-day constraints on DM, it is possible that it may not be WIMP-like and thermal in nature.
In the scenario proposed in this paper, we have focussed on the possible direct detection of high energy DM particles. Such particles cannot form the bulk of DM, which must be non-relativistic, but may be a small  population that lends itself to detection via methods different from those currently implemented at current DM detectors. One possible way such a component could exisit  at and around a specific high energy, would be due to its creation  by the decay of another significantly more massive non-thermal DM relic.
If the lighter DM particle interacts with nucleons, its cross-section at high energies may be detectable as neutrino-like cascades in a massive detector like IC. Using the neutrino-nucleon NC deep inelastic cross-section as a guiding analogy, we have applied this to the cluster of three $\sim$ PeV events seen at IC. 

Thus, this cluster of three events has a different origin from the remainder of the IC event sample, which we assume to be primarily astrophysical extra-galactic neutrinos. It results in a softer astrophysical spectral best-fit  than the one which includes the full-event sample. In this picture, the gap currently seen in the data between 400 TeV--1 PeV is physical, and the result of two distinct spectra. While it may partially get filled in or otherwise modified due to future data, it would remain as a demarcating feature between 2 fluxes of different origins, a UHE neutrino flux with a softer than currently estimated spectrum, and a DM flux that generates cascade interactions in the detector.
Additionally, the PeV events should continue to cluster in the 1--3 PeV region, with a galactic bias \cite{1311.5864} due to the fact that about half of the DM induced  PeV flux contribution is expected to be galactic. We note that at present 2 of the 3 events appear to come from the direction of the galaxy.
This scenario also provides a natural explanation for the lack of events beyond 3 PeV.
Other recent proposals, in addition to certain models of astrophysical sources referred to previously, which also account for the cut-off at PeV energies are discussed in \cite{Ema:2013nda, 1303.7320, 1308.1105, 1404.0622, 1404.2288, 1404.7025, 1407.0739}.

It is also to be noted that DM induced events will for the most part not contain energetic muon tracks, and will mostly be cascade-like. Thus, over time, if the IC sample contains a mixture of such events along with an astrophysical neutrino event component, the overall data will manifest a deficit in the ratio  of muon track to cascades  compared to the standard IC expectation of $1: 3$. 

 Additionally, for DM events in the 1--3 PeV range, some  extra-galactic contribution of cascades could come from the Northern hemisphere, because the lower DM-matter cross-section does not cause their flux to attenuate significantly in the earth at PeV energies, unlike neutrinos.
 These predictions  separate the present scenario from other DM induced indirect detection proposals \cite{1303.7320, 1308.1105}), and can be tested as IC gathers more data.

In conclusion, we have studied the possibility of detecting DM using large neutrino detectors,  via a relativistic and high energy component that may exist
in addition to the bulk of non-relativistic DM. As a specific example of the concept, we have applied it to recent events reported by IC, and also pointed out testable features of the scenario which can be used, with future data, to rule it out.

\begin{acknowledgments}
The authors would like to thank Nathan Whitehorn for his patient answering of many questions on the IC data and Arindam Chatterjee and Satyanarayan Mukhopadhyay for useful discussions related to this work.
AB is grateful to Tyce De Young for very insightful discussions and suggestions.
RG thanks Alejandro Ibarra for very useful discussions.
RG also acknowledges support from Fermilab via an Intensity Frontier Fellowship and thanks CETUP (Center for Theoretical Underground Physics and Related Areas) for partial support and hospitality during the 2014 Summer program.
RG and AG acknowledge support from a XII Plan DAE Neutrino Physics and Astrophysics Grant.
AG is also deeply appreciative of help from Mehedi Masud and Titas Chanda related to some of the relevant computational work.
This work was supported in part by the US Department of Energy contracts DE-FG02-04ER41298 and DE-
FG02-13ER41976 for AB.
\end{acknowledgments}

\bibliographystyle{JHEP}
\bibliography{DM-paper0}

\providecommand{\href}[2]{#2}\begingroup\raggedright\begin{thebibliography}{10}

\bibitem{Ibarra}
{A.Ibarra, Talk at Neutrino 2014}, {\it {Neutrinos and Dark Matter}}, .

\bibitem{1406.5200}
D.~Cline, {\it {A Brief Status of the Direct Search for WIMP Dark Matter}},
  \href{http://arxiv.org/abs/1406.5200}{{\tt arXiv:1406.5200}}.

\bibitem{1310.8642}
A.~Kusenko and L.~J. Rosenberg, {\it {Working Group Report: Non-WIMP Dark
  Matter}},  \href{http://arxiv.org/abs/1310.8642}{{\tt arXiv:1310.8642}}.

\bibitem{Nollett:2014lwa}
K.~M. Nollett and G.~Steigman, {\it {BBN And The CMB Constrain Neutrino Coupled
  Light WIMPs}},  \href{http://arxiv.org/abs/1411.6005}{{\tt arXiv:1411.6005}}.

\bibitem{Ade:2013zuv}
{\bf Planck} Collaboration, P.~Ade et~al., {\it {Planck 2013 results. XVI.
  Cosmological parameters}},  {\em Astron.Astrophys.} {\bf 571} (2014) A16,
  [\href{http://arxiv.org/abs/1303.5076}{{\tt arXiv:1303.5076}}].

\bibitem{Hooper:2011aj}
D.~Hooper, F.~S. Queiroz, and N.~Y. Gnedin, {\it {Non-Thermal Dark Matter
  Mimicking An Additional Neutrino Species In The Early Universe}},  {\em
  Phys.Rev.} {\bf D85} (2012) 063513,
  [\href{http://arxiv.org/abs/1111.6599}{{\tt arXiv:1111.6599}}].

\bibitem{Freedman:1973yd}
D.~Z. Freedman, {\it {Coherent neutrino nucleus scattering as a probe of the
  weak neutral current}},  {\em Phys.Rev.} {\bf D9} (1974) 1389--1392.

\bibitem{Billard:2013qya}
J.~Billard, L.~Strigari, and E.~Figueroa-Feliciano, {\it {Implication of
  neutrino backgrounds on the reach of next generation dark matter direct
  detection experiments}},  {\em Phys.Rev.} {\bf D89} (2014) 023524,
  [\href{http://arxiv.org/abs/1307.5458}{{\tt arXiv:1307.5458}}].

\bibitem{PhysRevLett.64.615}
{K.\ Griest and M.\ Kamionkowski}, {\it Unitarity limits on the mass and radius
  of dark-matter particles},  {\em Phys. Rev. Lett.} {\bf 64} (Feb, 1990)
  615--618.

\bibitem{Gandhi:1995tf}
R.~Gandhi, C.~Quigg, M.~H. Reno, and I.~Sarcevic, {\it {Ultrahigh-energy
  neutrino interactions}},  {\em Astropart.Phys.} {\bf 5} (1996) 81--110,
  [\href{http://arxiv.org/abs/hep-ph/9512364}{{\tt hep-ph/9512364}}].

\bibitem{Gandhi:1998ri}
R.~Gandhi, C.~Quigg, M.~H. Reno, and I.~Sarcevic, {\it {Neutrino interactions
  at ultrahigh-energies}},  {\em Phys.Rev.} {\bf D58} (1998) 093009,
  [\href{http://arxiv.org/abs/hep-ph/9807264}{{\tt hep-ph/9807264}}].

\bibitem{ab_in_progress}
Atri Bhattacharya et.al. in preparation.

\bibitem{Murase:2012xs}
K.~Murase and J.~F. Beacom, {\it {Constraining Very Heavy Dark Matter Using
  Diffuse Backgrounds of Neutrinos and Cascaded Gamma Rays}},  {\em JCAP} {\bf
  1210} (2012) 043, [\href{http://arxiv.org/abs/1206.2595}{{\tt
  arXiv:1206.2595}}].

\bibitem{Rott:2014kfa}
C.~Rott, K.~Kohri, and S.~C. Park, {\it {Superheavy dark matter and IceCube
  neutrino signals:bounds on decaying dark matter}},
  \href{http://arxiv.org/abs/1408.4575}{{\tt arXiv:1408.4575}}.

\bibitem{astro-ph/0403164}
K.~Ichiki, M.~Oguri, and K.~Takahashi, {\it {WMAP constraints on decaying cold
  dark matter}},  {\em Phys.Rev.Lett.} {\bf 93} (2004) 071302,
  [\href{http://arxiv.org/abs/astro-ph/0403164}{{\tt astro-ph/0403164}}].

\bibitem{Dev:2013yza}
P.~Bhupal~Dev, A.~Mazumdar, and S.~Qutub, {\it {Constraining Non-thermal and
  Thermal properties of Dark Matter}},  {\em Physics} {\bf 2} (2014) 26,
  [\href{http://arxiv.org/abs/1311.5297}{{\tt arXiv:1311.5297}}].

\bibitem{DelPopolo:2008mr}
A.~Del~Popolo, {\it {Dark matter and structure formation a review}},  {\em
  Astron.Rep.} {\bf 51} (2007) 169--196,
  [\href{http://arxiv.org/abs/0801.1091}{{\tt arXiv:0801.1091}}].

\bibitem{Esmaili:2012us}
A.~Esmaili, A.~Ibarra, and O.~L. Peres, {\it {Probing the stability of
  superheavy dark matter particles with high-energy neutrinos}},  {\em JCAP}
  {\bf 1211} (2012) 034, [\href{http://arxiv.org/abs/1205.5281}{{\tt
  arXiv:1205.5281}}].

\bibitem{1311.5864}
Y.~Bai, R.~Lu, and J.~Salvado, {\it {Geometric Compatibility of IceCube TeV-PeV
  Neutrino Excess and its Galactic Dark Matter Origin}},
  \href{http://arxiv.org/abs/1311.5864}{{\tt arXiv:1311.5864}}.

\bibitem{Alves:2013tqa}
A.~Alves, S.~Profumo, and F.~S. Queiroz, {\it {The dark $Z^{'}$ portal: direct,
  indirect and collider searches}},  {\em JHEP} {\bf 1404} (2014) 063,
  [\href{http://arxiv.org/abs/1312.5281}{{\tt arXiv:1312.5281}}].

\bibitem{Hooper:2014fda}
D.~Hooper, {\it {Z' Mediated Dark Matter Models for the Galactic Center
  Gamma-Ray Excess}},  \href{http://arxiv.org/abs/1411.4079}{{\tt
  arXiv:1411.4079}}.

\bibitem{Lai:2010vv}
H.-L. Lai, M.~Guzzi, J.~Huston, Z.~Li, P.~M. Nadolsky, et~al., {\it {New parton
  distributions for collider physics}},  {\em Phys.Rev.} {\bf D82} (2010)
  074024, [\href{http://arxiv.org/abs/1007.2241}{{\tt arXiv:1007.2241}}].

\bibitem{Buckley:2011vc}
M.~R. Buckley, D.~Hooper, J.~Kopp, and E.~Neil, {\it {Light Z' Bosons at the
  Tevatron}},  {\em Phys.Rev.} {\bf D83} (2011) 115013,
  [\href{http://arxiv.org/abs/1103.6035}{{\tt arXiv:1103.6035}}].

\bibitem{Kachelriess:2009zy}
M.~Kachelriess, P.~Serpico, and M.~A. Solberg, {\it {On the role of electroweak
  bremsstrahlung for indirect dark matter signatures}},  {\em Phys.Rev.} {\bf
  D80} (2009) 123533, [\href{http://arxiv.org/abs/0911.0001}{{\tt
  arXiv:0911.0001}}].

\bibitem{Aartsen:2013jdh}
{\bf IceCube} Collaboration, M.~Aartsen et~al., {\it {Evidence for High-Energy
  Extraterrestrial Neutrinos at the IceCube Detector}},  {\em Science} {\bf
  342} (2013), no.~6161 1242856, [\href{http://arxiv.org/abs/1311.5238}{{\tt
  arXiv:1311.5238}}].

\bibitem{Aartsen:2014gkd}
{\bf IceCube} Collaboration, M.~Aartsen et~al., {\it {Observation of
  High-Energy Astrophysical Neutrinos in Three Years of IceCube Data}},
  \href{http://arxiv.org/abs/1405.5303}{{\tt arXiv:1405.5303}}.

\bibitem{Adrian-Martinez:2014wzf}
{\bf ANTARES} Collaboration, S.~Adrian-Martinez et~al., {\it {Searches for
  Point-like and extended neutrino sources close to the Galactic Centre using
  the ANTARES neutrino Telescope}},  {\em Astrophys.J.} {\bf 786} (2014) L5,
  [\href{http://arxiv.org/abs/1402.6182}{{\tt arXiv:1402.6182}}].

\bibitem{Scherini:2014gja}
V.~Scherini, {\it {Updated results on Ultra-High Energy Neutrinos with the
  Pierre Auger Observatory}},  {\em PoS} {\bf Neutel2013} (2014) 058.

\bibitem{Margiotta:2014eaa}
{\bf KM3NeT} Collaboration, A.~Margiotta, {\it {Status of the KM3NeT project}},
   {\em JINST} {\bf 9} (2014) C04020.

\bibitem{Beacom:2004jb}
J.~F. Beacom and J.~Candia, {\it {Shower power: Isolating the prompt
  atmospheric neutrino flux using electron neutrinos}},  {\em JCAP} {\bf 0411}
  (2004) 009, [\href{http://arxiv.org/abs/hep-ph/0409046}{{\tt
  hep-ph/0409046}}].

\bibitem{Glashow:1960zz}
S.~L. Glashow, {\it {Resonant Scattering of Antineutrinos}},  {\em Phys.Rev.}
  {\bf 118} (1960) 316--317.

\bibitem{Bhattacharya:2011qu}
A.~Bhattacharya, R.~Gandhi, W.~Rodejohann, and A.~Watanabe, {\it {The Glashow
  resonance at IceCube: signatures, event rates and $pp$ vs. $p\gamma$
  interactions}},  {\em JCAP} {\bf 1110} (2011) 017,
  [\href{http://arxiv.org/abs/1108.3163}{{\tt arXiv:1108.3163}}].

\bibitem{Barger:2014iua}
V.~Barger, L.~Fu, J.~Learned, D.~Marfatia, S.~Pakvasa, et~al., {\it {Glashow
  resonance as a window into cosmic neutrino sources}},
  \href{http://arxiv.org/abs/1407.3255}{{\tt arXiv:1407.3255}}.

\bibitem{1403.3206}
A.~M. Taylor, S.~Gabici, and F.~Aharonian, {\it {A Galactic Halo Origin of the
  Neutrinos Detected by IceCube}},  \href{http://arxiv.org/abs/1403.3206}{{\tt
  arXiv:1403.3206}}.

\bibitem{1309.4077}
M.~Ahlers and K.~Murase, {\it {Probing the Galactic Origin of the IceCube
  Excess with Gamma-Rays}},  \href{http://arxiv.org/abs/1309.4077}{{\tt
  arXiv:1309.4077}}.

\bibitem{1309.2756}
S.~Razzaque, {\it {The Galactic Center Origin of a Subset of IceCube Neutrino
  Events}},  {\em Phys.Rev.} {\bf D88} (2013) 081302,
  [\href{http://arxiv.org/abs/1309.2756}{{\tt arXiv:1309.2756}}].

\bibitem{1311.7188}
C.~Lunardini, S.~Razzaque, K.~T. Theodoseau, and L.~Yang, {\it {Neutrino Events
  at IceCube and the Fermi Bubbles}},
  \href{http://arxiv.org/abs/1311.7188}{{\tt arXiv:1311.7188}}.

\bibitem{1405.3797}
M.~Kachelriess and S.~Ostapchenko, {\it {Neutrino yield from Galactic cosmic
  rays}},  \href{http://arxiv.org/abs/1405.3797}{{\tt arXiv:1405.3797}}.

\bibitem{1305.6606}
D.~Fox, K.~Kashiyama, and P.~Mészarós, {\it {Sub-PeV Neutrinos from TeV
  Unidentified Sources in the Galaxy}},  {\em Astrophys.J.} {\bf 774} (2013)
  74, [\href{http://arxiv.org/abs/1305.6606}{{\tt arXiv:1305.6606}}].

\bibitem{1310.7194}
M.~Gonzalez-Garcia, F.~Halzen, and V.~Niro, {\it {Reevaluation of the Prospect
  of Observing Neutrinos from Galactic Sources in the Light of Recent Results
  in Gamma Ray and Neutrino Astronomy}},  {\em Astropart.Phys.} {\bf 57-58}
  (2014) 39--48, [\href{http://arxiv.org/abs/1310.7194}{{\tt
  arXiv:1310.7194}}].

\bibitem{astro-ph/9609048}
V.~Berezinsky, P.~Blasi, and V.~Ptuskin, {\it {Clusters of galaxies as a
  storage room for cosmic rays}},  {\em Astrophys J.} {\bf 487} (1997)
  529--535, [\href{http://arxiv.org/abs/astro-ph/9609048}{{\tt
  astro-ph/9609048}}].

\bibitem{astro-ph/0601695}
A.~Loeb and E.~Waxman, {\it {The Cumulative background of high energy neutrinos
  from starburst galaxies}},  {\em JCAP} {\bf 0605} (2006) 003,
  [\href{http://arxiv.org/abs/astro-ph/0601695}{{\tt astro-ph/0601695}}].

\bibitem{1306.3417}
K.~Murase, M.~Ahlers, and B.~C. Lacki, {\it {Testing the Hadronuclear Origin of
  PeV Neutrinos Observed with IceCube}},  {\em Phys.Rev.} {\bf D88} (2013),
  no.~12 121301, [\href{http://arxiv.org/abs/1306.3417}{{\tt
  arXiv:1306.3417}}].

\bibitem{1303.1253}
H.-N. He, T.~Wang, Y.-Z. Fan, S.-M. Liu, and D.-M. Wei, {\it {Diffuse PeV
  neutrino emission from ultraluminous infrared galaxies}},  {\em Phys.Rev.}
  {\bf D87} (2013), no.~6 063011, [\href{http://arxiv.org/abs/1303.1253}{{\tt
  arXiv:1303.1253}}].

\bibitem{PhysRevLett.66.2697}
F.~Stecker, C.~Done, M.~Salamon, and P.~Sommers, {\it {High-energy neutrinos
  from active galactic nuclei}},  {\em Phys.Rev.Lett.} {\bf 66} (1991)
  2697--2700.

\bibitem{1305.7404}
F.~W. Stecker, {\it {PeV neutrinos observed by IceCube from cores of active
  galactic nuclei}},  {\em Phys.Rev.} {\bf D88} (2013), no.~4 047301,
  [\href{http://arxiv.org/abs/1305.7404}{{\tt arXiv:1305.7404}}].

\bibitem{astro-ph/9701231}
E.~Waxman and J.~N. Bahcall, {\it {High-energy neutrinos from cosmological
  gamma-ray burst fireballs}},  {\em Phys.Rev.Lett.} {\bf 78} (1997)
  2292--2295, [\href{http://arxiv.org/abs/astro-ph/9701231}{{\tt
  astro-ph/9701231}}].

\bibitem{1306.2274}
K.~Murase and K.~Ioka, {\it {TeV–PeV Neutrinos from Low-Power Gamma-Ray Burst
  Jets inside Stars}},  {\em Phys.Rev.Lett.} {\bf 111} (2013), no.~12 121102,
  [\href{http://arxiv.org/abs/1306.2274}{{\tt arXiv:1306.2274}}].

\bibitem{Murase:2008yt}
K.~Murase, S.~Inoue, and S.~Nagataki, {\it {Cosmic Rays Above the Second Knee
  from Clusters of Galaxies and Associated High-Energy Neutrino Emission}},
  {\em Astrophys.J.} {\bf 689} (2008) L105,
  [\href{http://arxiv.org/abs/0805.0104}{{\tt arXiv:0805.0104}}].

\bibitem{Loeb:2006tw}
A.~Loeb and E.~Waxman, {\it {The Cumulative background of high energy neutrinos
  from starburst galaxies}},  {\em JCAP} {\bf 0605} (2006) 003,
  [\href{http://arxiv.org/abs/astro-ph/0601695}{{\tt astro-ph/0601695}}].

\bibitem{1303.7320}
B.~Feldstein, A.~Kusenko, S.~Matsumoto, and T.~T. Yanagida, {\it {Neutrinos at
  IceCube from Heavy Decaying Dark Matter}},  {\em Phys.Rev.} {\bf D88} (2013),
  no.~1 015004, [\href{http://arxiv.org/abs/1303.7320}{{\tt arXiv:1303.7320}}].

\bibitem{1308.1105}
A.~Esmaili and P.~D. Serpico, {\it {Are IceCube neutrinos unveiling PeV-scale
  decaying dark matter?}},  {\em JCAP} {\bf 1311} (2013) 054,
  [\href{http://arxiv.org/abs/1308.1105}{{\tt arXiv:1308.1105}}].

\bibitem{Zavala:2014dla}
J.~Zavala, {\it {Galactic PeV neutrinos from dark matter annihilation}},
  \href{http://arxiv.org/abs/1404.2932}{{\tt arXiv:1404.2932}}.

\bibitem{Bhattacharya:2014vwa}
A.~Bhattacharya, M.~H. Reno, and I.~Sarcevic, {\it {Reconciling neutrino flux
  from heavy dark matter decay and recent events at IceCube}},  {\em JHEP} {\bf
  1406} (2014) 110, [\href{http://arxiv.org/abs/1403.1862}{{\tt
  arXiv:1403.1862}}].

\bibitem{Audren:2014bca}
B.~Audren, J.~Lesgourgues, G.~Mangano, P.~D. Serpico, and T.~Tram, {\it
  {Strongest model-independent bound on the lifetime of Dark Matter}},
  \href{http://arxiv.org/abs/1407.2418}{{\tt arXiv:1407.2418}}.

\bibitem{Ahlers:2014ioa}
M.~Ahlers and F.~Halzen, {\it {Pinpointing Extragalactic Neutrino Sources in
  Light of Recent IceCube Observations}},
  \href{http://arxiv.org/abs/1406.2160}{{\tt arXiv:1406.2160}}.

\bibitem{becker2005source}
J.~K. Becker, P.~L. Biermann, and W.~Rhode, {\it A source property based
  estimate of the neutrino flux from blazars and steep spectrum sources},  in
  {\em International Cosmic Ray Conference}, vol.~5, p.~9, 2005.

\bibitem{Gondolo:1995fq}
P.~Gondolo, G.~Ingelman, and M.~Thunman, {\it {Charm production and high-energy
  atmospheric muon and neutrino fluxes}},  {\em Astropart.Phys.} {\bf 5} (1996)
  309--332, [\href{http://arxiv.org/abs/hep-ph/9505417}{{\tt hep-ph/9505417}}].

\bibitem{Ema:2013nda}
Y.~Ema, R.~Jinno, and T.~Moroi, {\it {Cosmic-Ray Neutrinos from the Decay of
  Long-Lived Particle and the Recent IceCube Result}},  {\em Phys.Lett.} {\bf
  B733} (2014) 120--125, [\href{http://arxiv.org/abs/1312.3501}{{\tt
  arXiv:1312.3501}}].

\bibitem{1404.0622}
L.~Anchordoqui, V.~Barger, H.~Goldberg, J.~Learned, D.~Marfatia, et~al., {\it
  {End of the cosmic neutrino energy spectrum}},
  \href{http://arxiv.org/abs/1404.0622}{{\tt arXiv:1404.0622}}.

\bibitem{1404.2288}
K.~C.~Y. Ng and J.~F. Beacom, {\it {Cosmic neutrino cascades from secret
  neutrino interactions}},  \href{http://arxiv.org/abs/1404.2288}{{\tt
  arXiv:1404.2288}}.

\bibitem{1404.7025}
F.~W. Stecker and S.~T. Scully, {\it {Propagation of Superluminal PeV IceCube
  Neutrinos: A High Energy Spectral Cutoff or New Constraints on Lorentz
  Invariance Violation}},  \href{http://arxiv.org/abs/1404.7025}{{\tt
  arXiv:1404.7025}}.

\bibitem{1407.0739}
J.~G. Learned and T.~J. Weiler, {\it {A Relational Argument for a $\sim$PeV
  Neutrino Energy Cutoff}},  \href{http://arxiv.org/abs/1407.0739}{{\tt
  arXiv:1407.0739}}.

\end{thebibliography}\endgroup

\end{document}